\documentclass[%
 aip,
 amsmath,amssymb,
 reprint,%
]{revtex4-1}

\usepackage{graphicx}
\usepackage{dcolumn}
\usepackage{bm}

\usepackage[utf8]{inputenc}
\usepackage[T1]{fontenc}
\usepackage{mathptmx}
\usepackage{etoolbox}
\usepackage{caption}
\usepackage{subfigure}
\usepackage{indentfirst}
\usepackage{color}

\makeatletter
\def\@email#1#2{%
 \endgroup
 \patchcmd{\titleblock@produce}
  {\frontmatter@RRAPformat}
  {\frontmatter@RRAPformat{\produce@RRAP{*#1\href{mailto:#2}{#2}}}\frontmatter@RRAPformat}
  {}{}
}%
\makeatother
\begin{document}

\preprint{AIP/123-QED}

\title{Realization of higher Precision in Interferometric-weak-value-based Small-tilt Measurement}
\author{ChaoXia Zhang}
\affiliation{ 
College of Physics and Electronic Engineering, Shanxi University, Taiyuan 030006, People’s Republic of China
}%
\affiliation{%
State Key Laboratory of Quantum Optics and Quantum Optics Devices, Taiyuan 030006, People’s Republic of China 
}%
\author{YongLang Lai}
\affiliation{ 
College of Physics and Electronic Engineering, Shanxi University, Taiyuan 030006, People’s Republic of China
}%
\affiliation{%
State Key Laboratory of Quantum Optics and Quantum Optics Devices, Taiyuan 030006, People’s Republic of China 
}%
\author{RongGuo Yang}
 \altaffiliation{Email:yrg@sxu.edu.cn.}
\affiliation{ 
College of Physics and Electronic Engineering, Shanxi University, Taiyuan 030006, People’s Republic of China
}%
\affiliation{%
State Key Laboratory of Quantum Optics and Quantum Optics Devices, Taiyuan 030006, People’s Republic of China 
}%
\affiliation{%
Collaborative Innovation Center of Extreme Optics, Shanxi University, Taiyuan 030006, People’s Republic of China  
}%
\author{Kui Liu}
\affiliation{ 
College of Physics and Electronic Engineering, Shanxi University, Taiyuan 030006, People’s Republic of China
}%
\affiliation{%
State Key Laboratory of Quantum Optics and Quantum Optics Devices, Taiyuan 030006, People’s Republic of China 
}%
\affiliation{%
Collaborative Innovation Center of Extreme Optics, Shanxi University, Taiyuan 030006, People’s Republic of China  
}%
\author{Jing Zhang}
\affiliation{ 
College of Physics and Electronic Engineering, Shanxi University,  Taiyuan 030006, People’s Republic of China
}%
\affiliation{%
State Key Laboratory of Quantum Optics and Quantum Optics Devices, Taiyuan 030006, People’s Republic of China 
}%
\affiliation{%
Collaborative Innovation Center of Extreme Optics, Shanxi University, Taiyuan 030006, People’s Republic of China  
}%

\author{HengXin Sun}
\affiliation{ 
College of Physics and Electronic Engineering, Shanxi University, Taiyuan 030006, People’s Republic of China
}%
\affiliation{%
State Key Laboratory of Quantum Optics and Quantum Optics Devices, Taiyuan 030006, People’s Republic of China 
}%
\affiliation{%
Collaborative Innovation Center of Extreme Optics, Shanxi University, Taiyuan 030006, People’s Republic of China  
}%

\author{ JiangRui Gao}
\affiliation{ 
College of Physics and Electronic Engineering, Shanxi University, Taiyuan 030006, People’s Republic of China
}%
\affiliation{%
State Key Laboratory of Quantum Optics and Quantum Optics Devices, Taiyuan 030006, People’s Republic of China 
}%
\affiliation{%
Collaborative Innovation Center of Extreme Optics, Shanxi University, Taiyuan 030006, People’s Republic of China  
}%

\date{\today}
\setlength{\parindent}{2em}
\begin{abstract}
We experimentally realize a great precision enhancement in the small tilt measurement by using a Sagnac interferometer and balanced homodyne detection (BHD) of high-order optical modes, together with the weak value amplification (WVA) technique. Smaller minimum measurable tilt (MMT) and higher signal-to-noise ratio (SNR) can be obtained by using BHD, compared with the split detection (SD). The precision of 3.8 nrad can be obtained under our present experimental condition. It is shown that combining WVA technique and BHD can strengthen each other’s advantages and can behave better for some special application scenarios, such as extremely weak output, wider measurement bandwidth, etc. Moreover, the precision can be further enhanced by experimental parameter optimization. 
\end{abstract}

\maketitle

\begin{quotation}
  Quantum measurement is not only an important tool to explore the micro-world, but also a key to the interpretation of quantum mechanics and the application of quantum information. For a long time, the improvement of precision measurement technology and the breakthrough of higher measurement precision are the goals that people strive to pursue. As one of the basic facets in precision measurement, optical tilt and displacement measurement has attracted more attentions in recent years, which has been applied in many fields, such as biological measurement\cite{taylor2013biological}, atomic force microscopy\cite{meyer1988novel}, high-resolution quantum imaging\cite{lugiato2002quantum, delaubert2008quantum,kolobov1999spatial}, and gravitational wave detection\cite{abbott2009ligo,scientific2017gw170104}.\\
 \indent The weak value amplification (WVA) technique, which was first proposed by Aharanov et al in 1988\cite{aharonov1988result} and widely used in many different systems, has been proved to have its advantages for precision enhancement not only theoretically\cite{Kedem2012,jordan2014technical} but also experimentally\cite{viza2015}. The interferometric WVA technique was used to amplify very small transverse deflections of an optical beam and WVA factors of over 100 are achieved\cite{dixon2009}. Results of a high precision small-displacement mesurement based on WVA technique, which demonstrate the capability of the WVA scheme to overcome the precision limit set by the saturation of the detectors\cite{xu2020approaching}. WVA in the azimuthal degree of freedom was demonstrated and a sensitive estimation of angular rotation of the spatial-mode light with the corresponding effective amplification factors as large as 100 was got\cite{magana2014}. The precision enhancement of weak-value-based metrology together with the power-recycling technique was investigated\cite{wang2016}. Moreover, WVA technique was also adopted on an integrated photonic platform and enhanced on-chip phase measurement with a 7 dB signal enhancement was achieved\cite{song2021}. It is worth noting that the detecting devices of almost all of the above WVA systems are usually based on the split detection (SD) or CCD detection.\\ 
 \indent On the other hand, high-order-mode light or spatial squeezed light can be used in high-precision displacement and tilt measurement due to their spatial transverse property. The precision was improved 1.5 times compared with the quantum noise limit (QNL) by using a spatial squeezed light of 3.3 dB\cite{treps2003quantum} in a small-displacement measurement. Our group demonstrated 1.4 times  precision enhancement with the $TE{{M}_{10}}$ mode compared with the $TE{{M}_{00}}$ mode\cite{sun2014small}and also realized 1.2 times precision improvement with a spatial squeezed light of 2.2 dB compared with the shot noise limit (SNL)\cite{sun2014experimental}. Recently, our group proposed an efficient method to generate high-order mode squeezing light and constructed a fourth-order tilt and displacement squeezed light for spatial tilt and displacement measurement, which improved the  precision by 3.2 times and 2.7 times compared with $TE{{M}_{00}}$ mode, respectively\cite{li2022higher}.\\
  \indent In addition, the detection system is also of great importance. The balanced homodyne detection (BHD), which usually consists of a pair of ETX-500 photodiodes (detection bandwidth:140 MHz, dark current: 12nA, NEP:125nW), allows us to place different carriers in the spectrum and to detect very weak signals, and can also be used in squeezed/entangled light observation, optical homodyne tomography, etc., to get the quantum noise. It is shown that the $TE{{M}_{10}}$ homodyning scheme outperforms SD for all values of squeezing although the QNL of displacement measurement can be surpassed by using squeezed light of appropriate spatial modes for both schemes\cite{hsu2004optimal} and it is proved that the standard SD is only $64\%$ efficient relative to the $TE{{M}_{10}}$ homodyne detection\cite{delaubert2006tem}.\\
\indent In this letter, we demonstrate the precision enhancement in an optical small tilt measurement applying WVA+BHD technique both theoretically and experimentally. The combination of WVA and BHD techniques based on the high order mode brings a better performance to detect extremely weak signal with fast detection speed and large frequency bandwidth.

\end{quotation}

\vspace{-0.4cm}
\section{\label{sec:level1}Theoretical Derivation }
\subsection{Weak Value Measurement Process}
The Sagnac interferometer containing the tilt modulation system (TMS), which is introduced into the signal field, is the main part of the weak value measurement system (WVMS). The weak-value measurement operation involves two quantum states, the system and the pointer. The path information of Sagnac interferometer is called the system, expressed by $\left| + \right\rangle$ for clockwise path and $\left| - \right\rangle$ for counterclockwise path\cite{howell2010interferometric}. We refer to the beam’s transverse amplitude distributions as the pointer, described by $\left| {{\psi }_{i}} \right\rangle$. A standard weak measurement process consists of three steps: preselection, weak interaction, and post-selection. Preselection couples the initial state of the system $\left| i \right\rangle$ and the pointer $\left| {{\psi }_{i}} \right\rangle$ in product state $\left| i \right\rangle \left| {{\psi }_{i}} \right\rangle$. Weak interaction produces entanglement between the system and the pointer states, and the interaction Hamiltonian is
\begin{equation}
{{H}_{\operatorname{int}}}=\lambda g(t)\hat{A}\text{}k\hat{x},
\end{equation}
where $g(t)$ is the normalized parameter which varies with time. $\lambda $ is the coupling coefficient (sufficiently small). $\hat{A}$ is the operator of the system state. $k$ is the transverse momentum kick of light beam induced by the mirror vibration and $\hat{x}$ is the operator of the pointer state. Post-selection process is the projection of the entangled state onto the final state $\left| f \right\rangle$ of the system. The final state is
\begin{equation}
\begin{split}
\left| {{\psi }_{f}} \right\rangle&={\left\langle  f \right|\hat{U}\left| i \right\rangle \left| {{\psi }_{i}} \right\rangle }/{\sqrt{{{P}_{m}}}}\;={\left\langle  f \right|\exp (-i\int{{{H}_{\operatorname{int}}}dt}\left| i \right\rangle \left| {{\psi }_{i}} \right\rangle }/{\sqrt{{{P}_{m}}}}\;\\
&\approx {\left\langle  f \right|(1-i\hat{A}k\hat{x}\left| i \right\rangle \left| {{\psi }_{i}} \right\rangle }/{\sqrt{{{P}_{m}}}}\;\text{=}{\left\langle  f | i \right\rangle (1-i{{A}_{w}}k\hat{x}\left| {{\psi }_{i}} \right\rangle }/{\sqrt{{{P}_{m}}}}\;\\
&=\left| {{\psi }_{i}}(x-{{A}_{w}}k) \right\rangle,
\end{split}
\end{equation}
where ${{A}_{w}}=\frac{\left\langle  f \right|A\left| i \right\rangle }{\left\langle  f | i \right\rangle }$ is the weak value. ${{P}_{m}}={{\left| \left\langle  f | i \right\rangle  \right|}^{2}}$ is the post-selection probability with initial state $\left| i \right\rangle =\frac{1}{\sqrt{2}}({{e}^{-i{\phi }/{2}\;}}\left| + \right\rangle +{{e}^{i{\phi }/{2}\;}}\left| - \right\rangle )$ and final state $\left| f \right\rangle =\frac{1}{\sqrt{2}}(\left| + \right\rangle -\left| - \right\rangle )$, where $\phi$  is the relative phase between the clockwise and the counterclockwise paths of the Sagnac interferometer\cite{howell2010interferometric}. ${{\psi }_{i}}$ denotes the transverse amplitude distribution of one-dimensional Hermite-Gauss (HG) modes. One can see that, there exist a transverse change of the beam given by ${{A}_{w}}k$. The smaller the relative phase $\phi$, the weaker the dark-port beam, which result in a larger ${{A}_{w}}$ and then an amplified transverse change is ${{A}_{w}}k$.

\subsection{\label{sec:level3} Balanced Homodyne Detection}
\vspace{-0.2cm}
\begin{figure}[htbp]
\centering
\subfigure[]{
\begin{minipage}[t]{0.5\linewidth}
\centering
\includegraphics[width=1.5in]{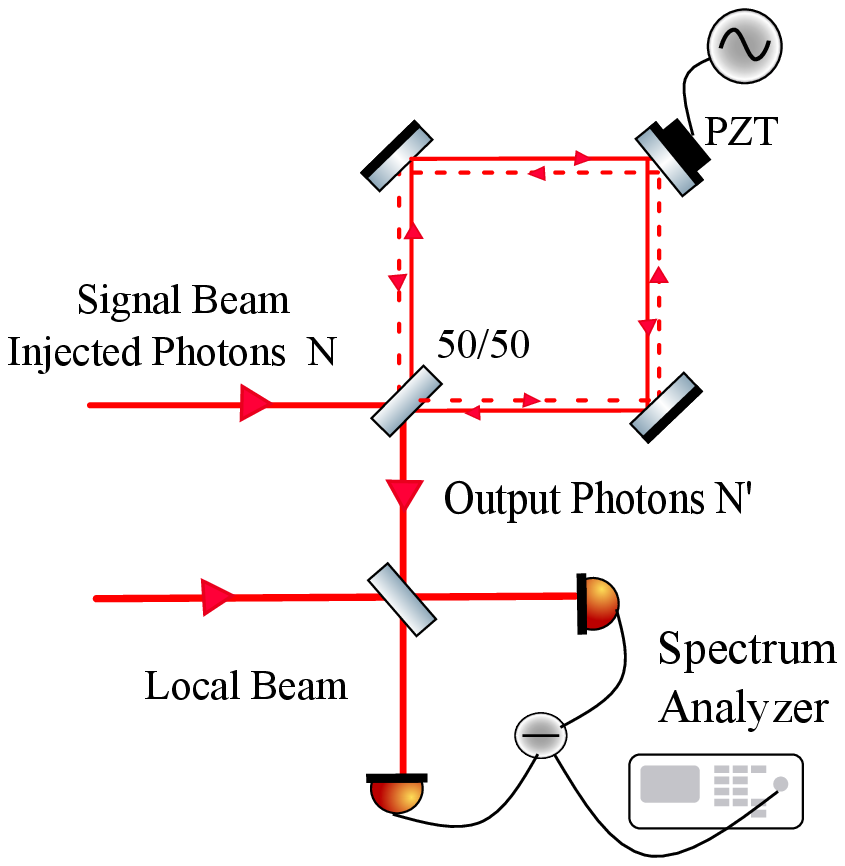}
\end{minipage}%
}%
\subfigure[]{
\begin{minipage}[t]{0.5\linewidth}
\centering
\includegraphics[width=1.5in]{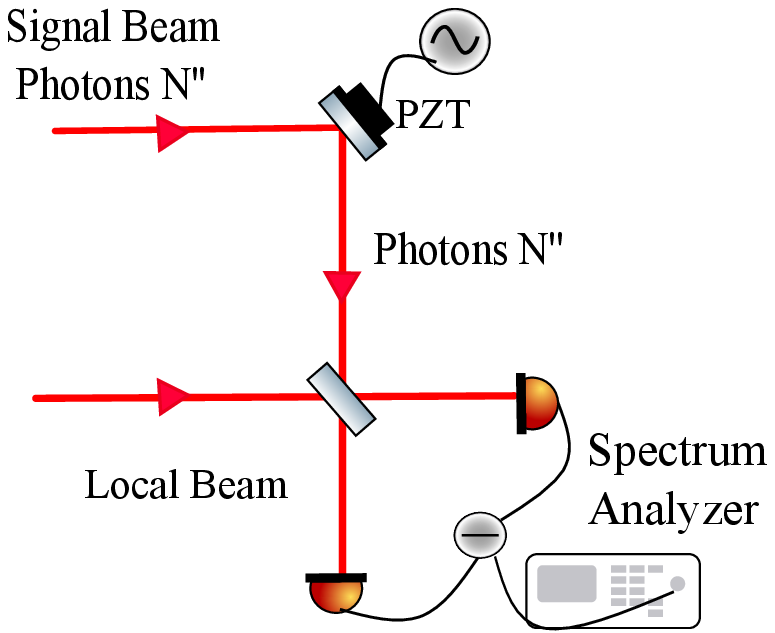}
\end{minipage}%
}%
\centering
\caption{BHD system of Weak Value Measurement (a) and Conventional Measurement (b). PZT: piezoelectric. transducers}
\end{figure}
A small transverse change ${{A}_{w}}k$ of a $TE{{M}_{00}}$ mode can induce many higher order modes, then the pointer state in one dimension has the form:
\begin{eqnarray}
\begin{split}
{{\psi }_{i}}(x-{{A}_{w}}k)&\approx {{\psi }_{0}}(x)-i\frac{{{\omega }_{o}}{{A}_{w}}k}{2}{{\psi }_{1}}(x)\\
&={{\psi }_{0}}(x)+i\frac{\pi {{\omega }_{o}}{{A}_{w}}\theta }{\lambda}{{\psi }_{1}}(x),
\end{split}
\end{eqnarray}
\begin{eqnarray}
{{\psi}_{n}}(x)={{\left( \frac{2}{\pi {{\omega }_{0}}} \right)}^{1/4}}\frac{1}{\sqrt{n!{{2}^{n}}}}{{H}_{n}}(\frac{\sqrt{2}x}{{{\omega }_{0}}}){{e}^{\frac{-{{x}^{2}}}{{{\omega }_{0}}^{2}}}}, n=0,1,2...,
\end{eqnarray}
where ${{\omega }_{0}}$ is the waist of the $TE{{M}_{00}}$ mode, and ${{H}_{n}}(x)$ is the Hermite polynomial with $n$ representing the order of the HG mode. Obviously, $TE{{M}_{10}}$ induced by the translation of $TE{{M}_{00}}$ contains the tilt information, which can be extracted by using $TE{{M}_{10}}$ mode as the local beam.\\
\indent Considering the BHD system, we can obtain
\begin{equation}
N_{-}^{BHD,n}=\sqrt{{{N}_{LO}}}(2\sqrt{{{N}^{'}}}{{A}_{w}}k{{\omega }_{0}}+\delta \hat{X}_{S}^{-}),
\end{equation}
 where the first term is the signal part including the small-tilt information and the second term is the quantum noise. ${{N}^{'}}$ and ${{N}_{LO}}$ are the mean photon numbers of the signal and local beams, respectively. $\overset{\wedge }{\mathop{\delta {{X}_{1}}}}\,$ is the quantum noise of the signal beam and $k$ is the transverse momentum kick of light beam. \\
 \indent Then the signal-to-noise (SNR) of the tilt measurement applying BHD system is
\begin{equation}
SNR={{\left( \frac{2\sqrt{{{N}^{'}}}{{A}_{w}}k{{\omega }_{0}}}{\delta X_{S}^{-}} \right)}^{2}},
\end{equation}
consider that ${{N}^{'}}=N{{\left| \left\langle  f | i \right\rangle  \right|}^{2}}$, in which $N$  is the number of photons injected into Sagnac interferometer (bright port). Then ${{N}^{'}} $ can also be regarded as the number of photons output from the dark port of the Sagnac interferometer and can be detected by the BHD system. For coherent light, the corresponding SNR can be defined as 
\begin{equation}
SNR\text{=}{{\left( 2\sqrt{N{{\left| \left\langle  f | i \right\rangle  \right|}^{2}}}{{A}_{w}}k {{\omega }_{0}} \right)}^{2}}={{\left( 2\sqrt{N}\cos \frac{\phi }{2}k {{\omega }_{0}} \right)}^{2}}
\end{equation}
\indent The minimum measurable tilt (MMT) can be generally defined by the tilt with $SNR\text{=}1$, which is given by
\begin{equation}
{{\theta }_{\min }}=\frac{\lambda }{4\pi {{\omega }_{0}}\sqrt{N}\cos \frac{\phi }{2}}
\end{equation}
\indent As is known, the SNR and the minimum measurable value is related with the measurement precision. The higher the SNR or the smaller the minimum measurable value, the higher the measurement precision. The initial and final states of the system tend to be orthogonal when the relative phase $\phi$ is small enough, meanwhile the SNR becomes higher and the minimum measurable value becomes smaller with the same waist and photon number. When $\phi =0$, the expression of the MMT corresponds to the standard quantum limit. In addition, when the dark port is tuned dark enough, the SNR and MMT are related to the injected photon number $N$ but independent of the detected signal photon number ${{N}^{'}}$, which means that effective beam tilt information can be obtained by post-selecting only extremely small part of those displaced photons. It is worth to mention that destructive interference is the essence of the WVA process. However, at the dark port, destructive interference only happens to the $TE{{M}_{00}}$ mode and simultaneously constructive interference to the $TE{{M}_{10}}$  mode without any loss of the tilt information. In this sense, the post-selection process is not simply a background filtering process, but a destructive interference process.\\
\begin{figure}[htbp]
\centering
\subfigure[]{
\begin{minipage}[t]{1\linewidth}
\hspace{-1.05cm}
\centering
\includegraphics[width=2.5in]{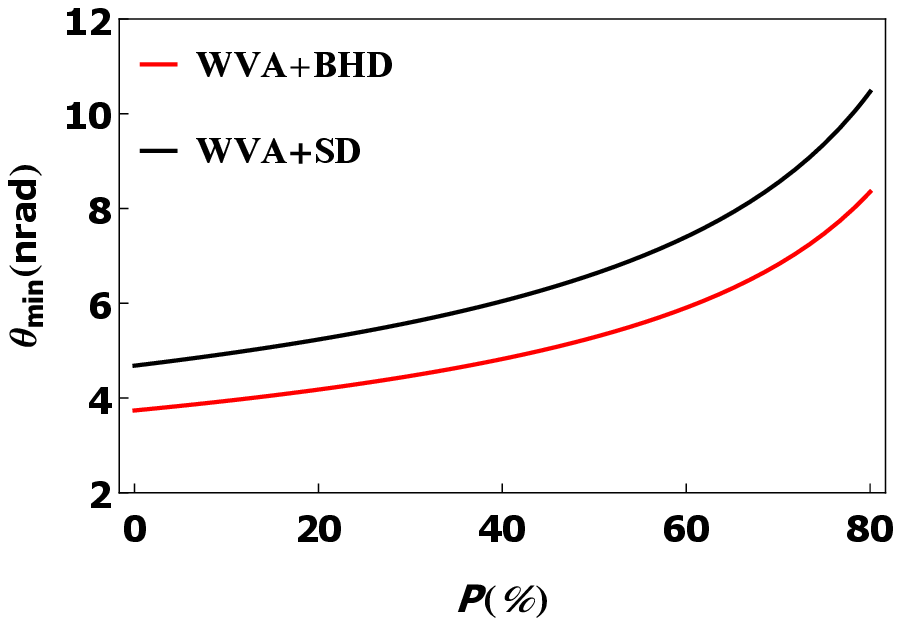}
\end{minipage}%
}%

\subfigure[]{
\begin{minipage}[t]{1.05\linewidth}
\hspace{-1.05cm}
\centering
\includegraphics[width=2.4in]{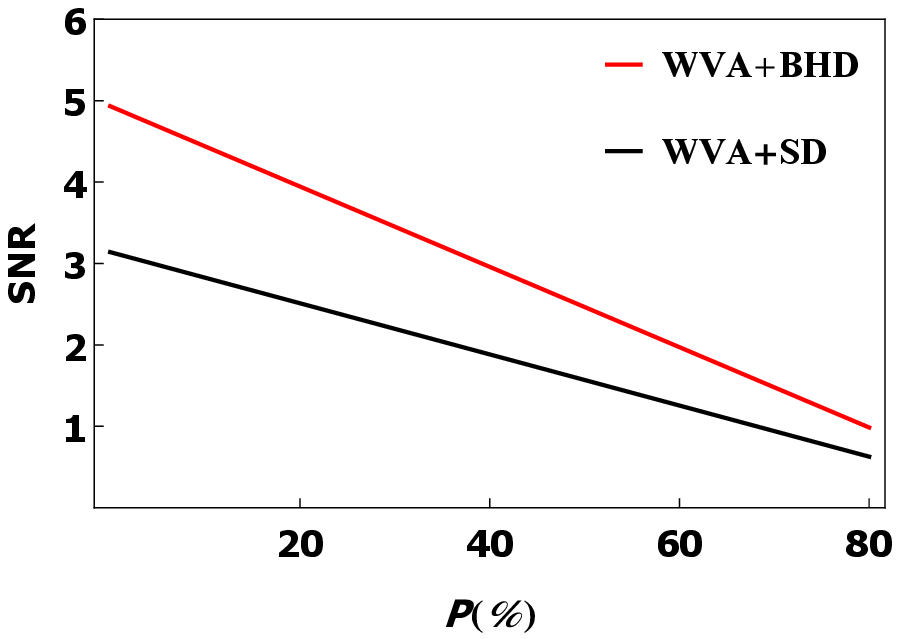}
\end{minipage}%
}%
\centering
\caption{(a) The MMT (a) and the SNR (b) versus the post-selection probability. $N=\text{5}\text{.35265}\times \text{1}{{\text{0}}^{\text{11}}}$, ${{\omega }_{0}}\text{=}60\mu m$.}
\end{figure}\\ 
\indent For comparation, the conventional BHD system \cite{sun2014small} is given in Fig.1(b) and  ${{N}^{''}}$ is the photon number detected from the BHD system. Obviously, in the conventional BHD system, ${{N}^{''}}$ is usually very small due to the power limitation of the local beam and the saturation of the detector. Thus the advantage of WVA technique is that $N$ can be much greater than ${{N}^{''}}$ under actual experimental conditions. Actually, the small-tilt information of all photons can be extracted by detecting a small number of photons, which solves the problem that the measurement  precision is limited by the saturation of the detector in the conventional scheme.
Since the injected photon number can be extremely larger than the detected photon number, the SNR can be improved greatly. Here the "WVA+BHD" system is more convenient to get better precision than our previous "High-order-mode+BHD" system \cite{sun2014small}, because increasing photon number is easier than increasing the order of the high-order modes.\\ 
\indent In addition, the SNR and the MMT detected by SD can be obtained as
\begin{equation}
SN{{R}_{SD}}=\frac{2}{\pi }{{\left( 2\sqrt{N}\cos \frac{\phi }{2}k{{\omega }_{0}} \right)}^{2}},
\end{equation}
\begin{equation}
\theta _{\min }^{SD}=\sqrt{\frac{\pi }{2}}\frac{\lambda }{4\pi {{\omega }_{0}}\sqrt{N}\cos \frac{\phi }{2}}
\end{equation}
\indent As is shown in fig.2, under the same conditions, the“WVA+BHD”scheme results in higher SNR and smaller MMT, compared with the“WVA+SD” scheme. Moreover, the “WVA+BHD” scheme can treat with lower power (several $\mu W$) and immune to the electronic noise level (ENL), due to the amplification of high-order-mode BHD local beam. The detection efficiency of BHD scheme is 36$\%$ higher than that of SD\cite{delaubert2006tem}.

\section{Experimental Measurement}
\vspace{0.16cm}
\begin{figure}[htbp]
\includegraphics[width=3in]{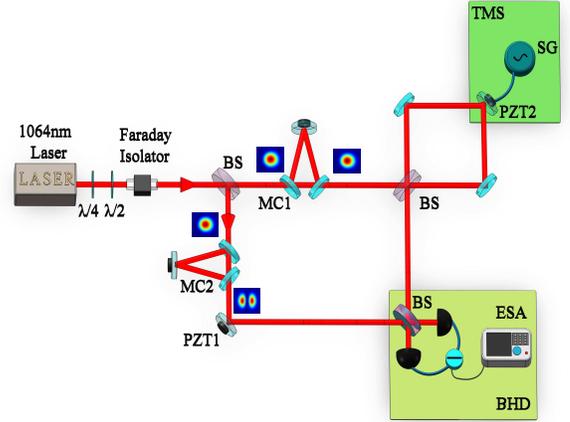}
\caption{\label{fig:epsart} Experimental setup. MC1: mode cleaner, MC2: mode converters, TMS: tilt modulation system, PZT1, PZT2: piezoelectric transducers, SG: signal generator, BS: 50/50 beam splitter, BHD: balanced homodyne detection, ESA: electronic spectrum analyzer.}
\end{figure}

\vspace{-0cm}
The experimental setup is shown in Fig.3. A continuous wave solid-state YAG laser operating at 1064 nm is used to drive the system. The laser beam can be divided into two beams. One beam passing through the mode-conversion cavity MC2 is called the local beam. While the other beam passing through the mode-cleaner cavity MC1 is the signal beam, which next enter the WVMS  and is tuned by the TMS. After that, the local and signal beams are coupled into the BHD system and the BHD output can be analyzed by an electronic spectrum analyzer (ESA).\\
\begin{figure}[htbp]
\subfigure{
\hspace{-1cm}
\begin{minipage}[t]{0.91\linewidth}
\centering
\includegraphics[width=3.3in]{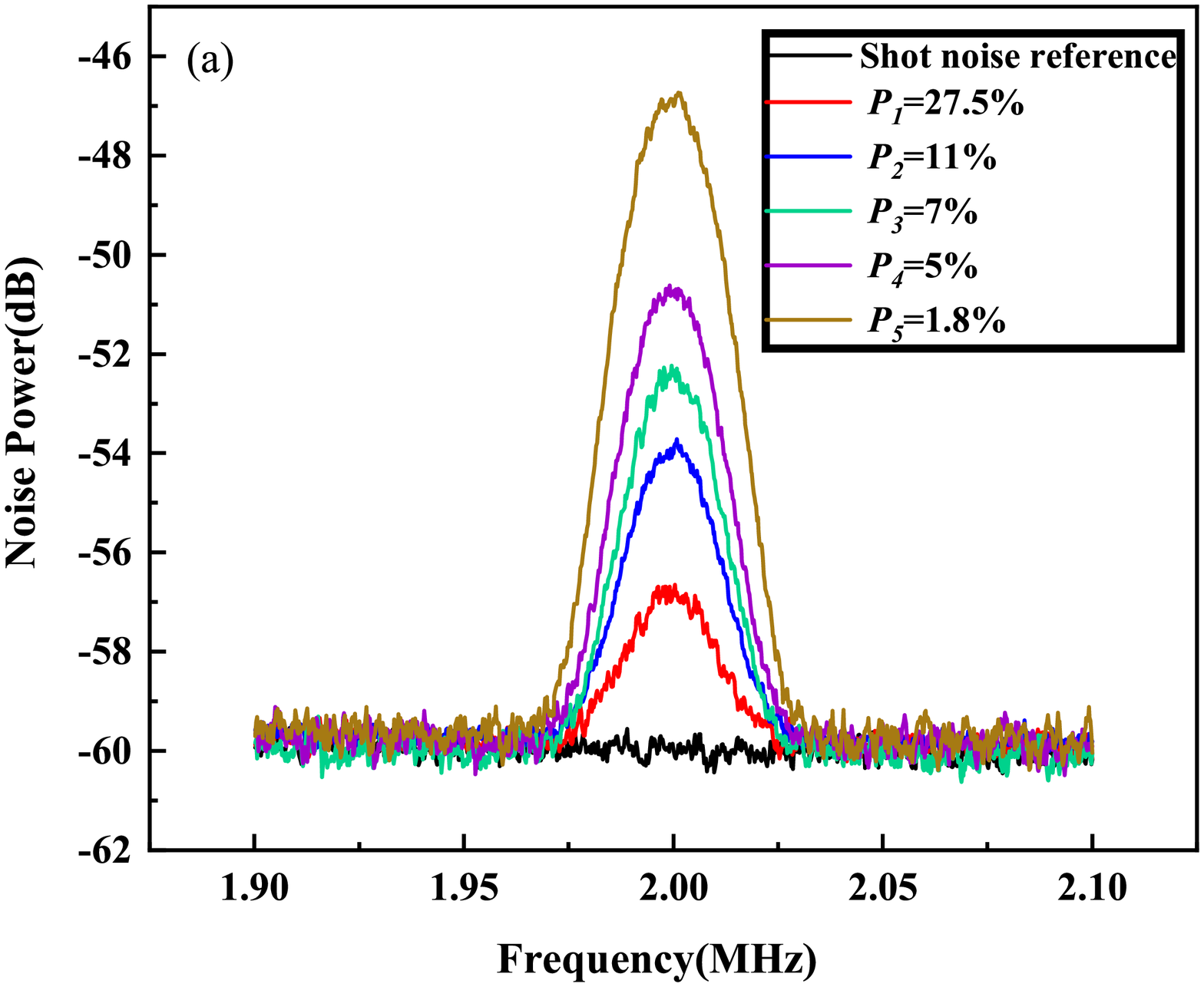}
\end{minipage}%
}%
\hspace{5cm}
\subfigure{
\hspace{-1cm}
\begin{minipage}[t]{0.9\linewidth}
\centering
\includegraphics[width=3.3in]{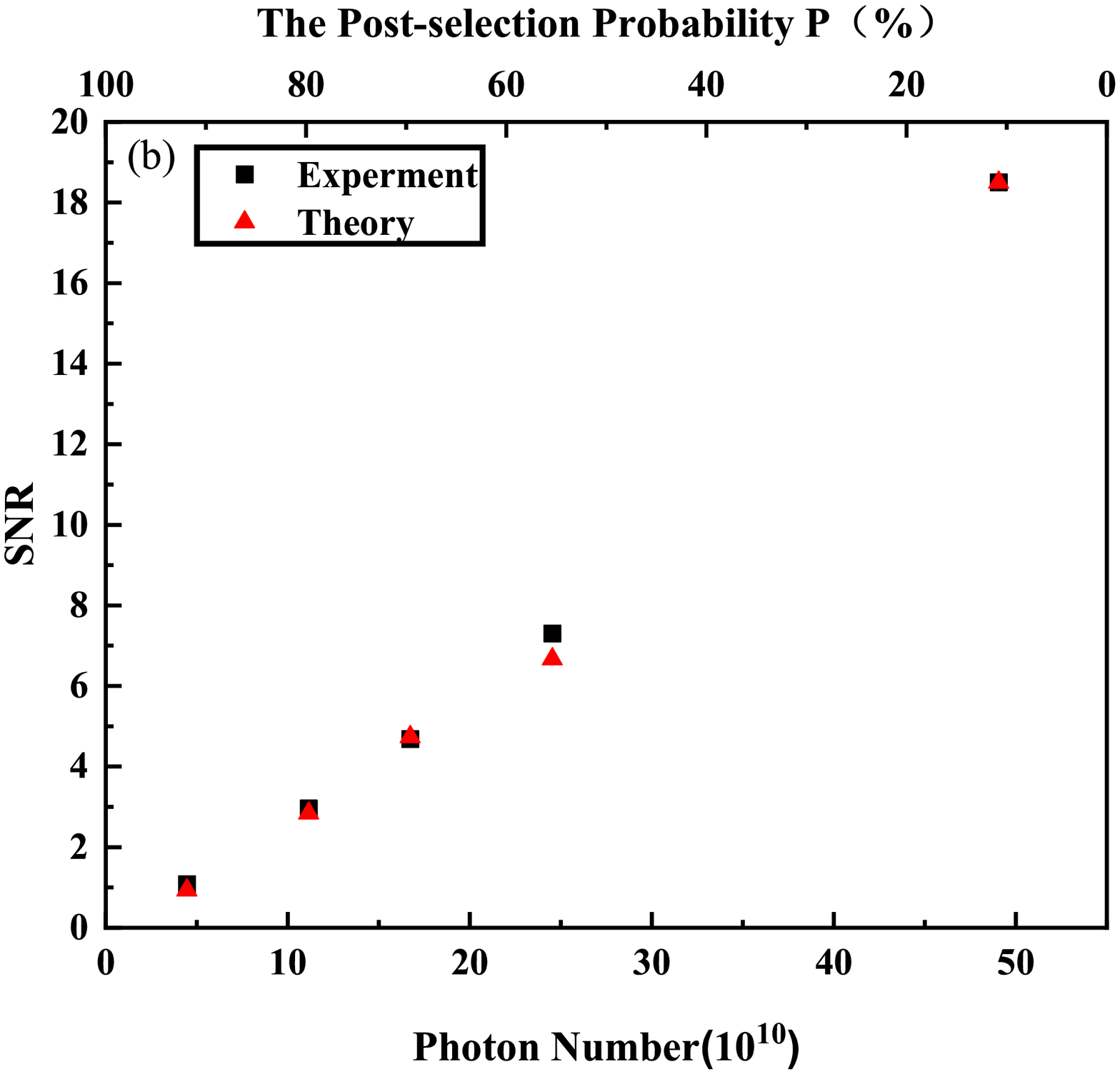}
\end{minipage}%
}%
\centering
\caption{(a)Noise power spectrum with different post-selection probabilities (b) SNR versus the injected photon numbers of the Sagnac interferometer/the post-selection probability.}
\end{figure}
\vspace{0cm}
\indent The input and output ports of the Sagnac interferometer are corresponding to the pre-selection and post-selection steps, respectively. While the weak interaction is from the TMS. The TMS consists of a mirror mounted on a piezoelectric transducer (PZT2) and a connected signal generator (SG). PZT2 is driven by a sine wave signal with a mechanical resonance frequency 2MHz . The experimental parameters are: $TE{{M}_{00}}$ waist ${{\omega }_{0}}\text{=}60\mu m$, the local beam power ${{P}_{local}}=1mW$, the resolution bandwidth $RBW=24kHz$, video bandwidth $VBW=130Hz$, and analyzing frequency $f=2MHz$.\\
\begin{figure}[htbp]
\subfigure{
\hspace{-1.0cm}
\begin{minipage}[t]{0.9\linewidth}
\centering
\includegraphics[width=3.3in]{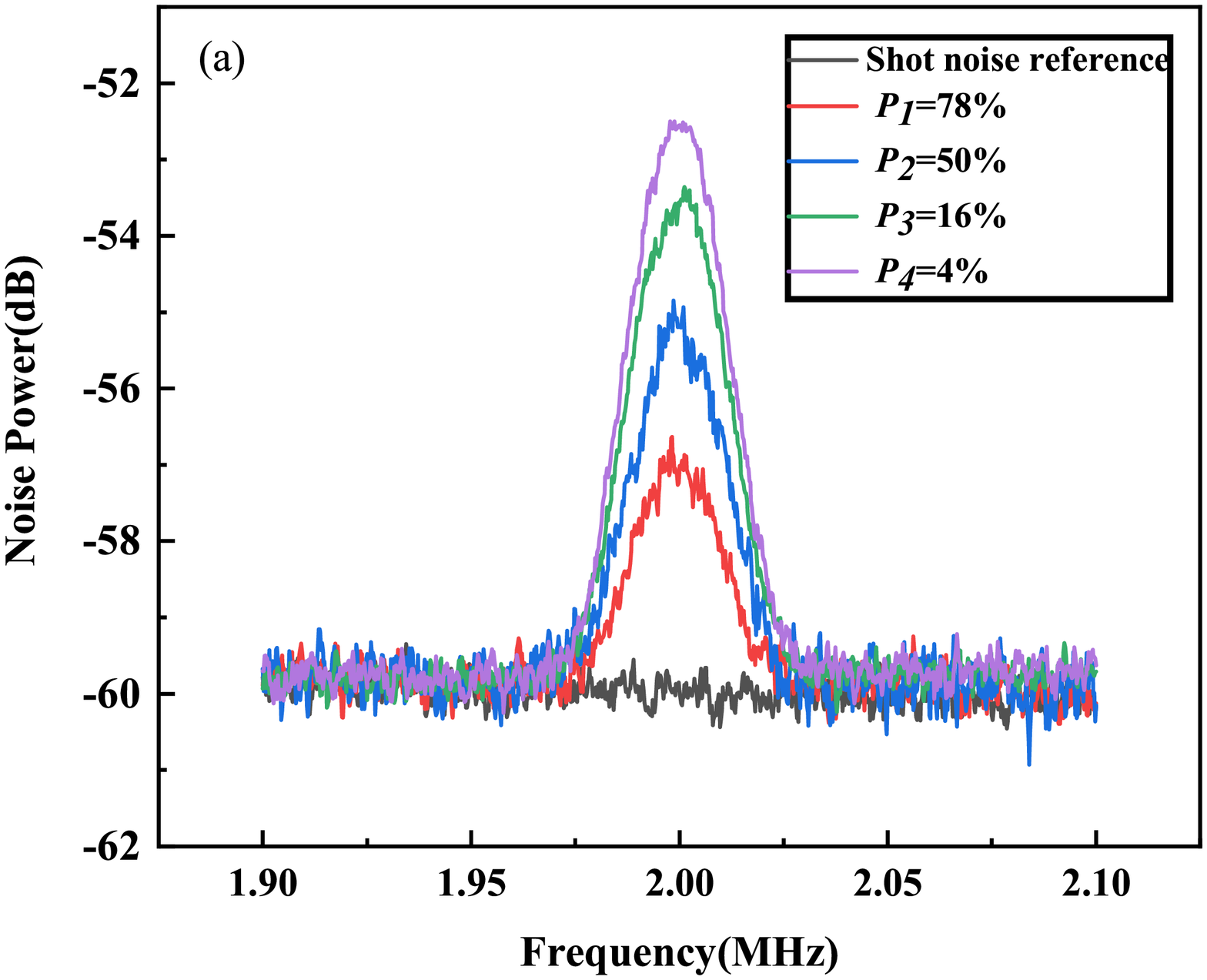}
\end{minipage}%
}%

\subfigure{
\hspace{-1.0cm}
\begin{minipage}[t]{0.9\linewidth}
\centering
\includegraphics[width=3.3in]{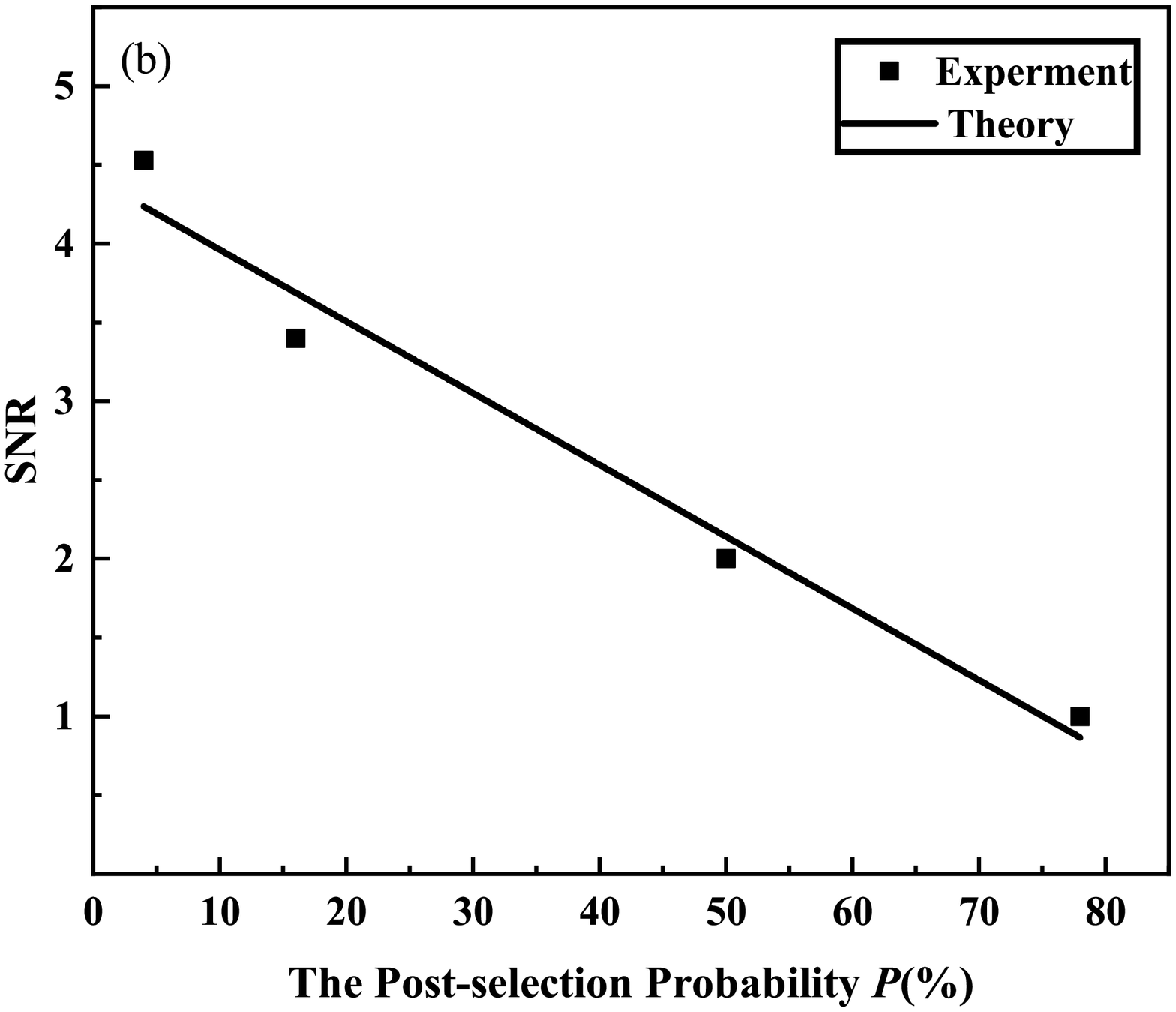}
\end{minipage}%
}%
\centering
\caption{Noise power spectrum (a) and the SNR (b) versus the post-selection probability. (The solid line is the theoretical fitting curve.)}
\end{figure}
\indent In Fig.4(a), the output power of the WVMS is fixed at $55\mu W$, and the injected optical power is chosen to be $200\mu W,500\mu W,800\mu W,1.1mW$ and $3.2mW$, respectively. In Fig 4(b), the tilt corresponding to the maximum SNR is regarded as a reference value, based on which the theoretical values of other points (represented by red triangles) can be derived. It is obvious that the signal beam power detected by the BHD is much smaller than the local beam power. During this process, the  piezo driving voltage of the TMS remains unchanged. The post-selection probability ${{P}_{m}}$ is equal to the ratio of the output power ${{P}_{out}}$ to the input power ${{P}_{in}}$, i.e., ${{P}_{m}}\text{=}{{\left| \left\langle  f | i \right\rangle  \right|}^{2}}\text{=}{{\sin }^{2}}\frac{\phi }{2}={{{P}_{out}}}/{{{P}_{in}}}\;,m=1,2,3....$  As is shown in Fig.4(a), the smaller the post-selection probability, the larger the noise power. Here the amplified signal measured by our system only depend on the injected beam power of the Sagnac interferometer (WVMS). As is shown in Fig.4(b), the lower the post-selection probability, the higher the SNR.\\

\indent In Fig.5, the injected beam power of the Sagnac interferometer is fixed at $70\mu W$ and the output beam power is tuned to be $55\mu W,35\mu W,11\mu W,4\mu W$, respectively. The  piezo driving voltage of the TMS still remains unchanged. Here, a low injected beam power is chosen, for better investigating the influence of changing the dark port (from dark to bright) on the measurement results.

It is important to note that the local beam power should still be much greater than the signal beam power when the dark port becomes bright.  As is shown in Fig.5, the smaller the post-selection probability, the larger the signal power and the SNR. With the same injected power of the Sagnac interferometer, the weaker the dark port, the smaller the post-selection probability, the more tilt information contained.\\ 
\vspace{-0.4cm}
\begin{table}[htbp]
\caption{\label{tab:table2}The relationship table among the input and output power, the piezo driving voltage, and the MMT. ${{P}_{m}}=\frac{{{P}_{out}}}{{{P}_{in}}}\approx \text{3}\text{.3 }\!\%\!\!\text{ }$, $SNR=1$.}
\begin{ruledtabular}
\begin{tabular}{cccccccc}
 ${{P}_{in}}$\footnotemark[1]($\mu W$)&${{P}_{out}}$\footnotemark[2]($\mu W$)&
 V\footnotemark[3](mV)&${{\theta }_{\min }}(nrad)$\\
\hline
210& 7 & 1000 & 8.29\\
300& 10 & 800 & 6.93\\
400& 13 & 715 & 6.01 \\
500& 17 & 660 & 5.37 \\
750& 25 & 560 & 4.39 \\
1000& 33 & 400 & 3.8  \\
\end{tabular}
\end{ruledtabular}
\footnotetext[1]{Sagnac interferometer injected optical power}
\footnotetext[2]{Sagnac interferometer output optical power}
\footnotetext[3]{Piezo Driving Voltage (mV).}
\end{table}\\
\indent In Table.1, the driving voltage on PZT2 is changed to keep SNR=1(the MMT, 3dB noise on the spectrometer) and the post-selection probability remains unchanged, for each measurement. 
 $RBW=10kHz$, for more convenient and precise. As is shown, when input and output power are $1000\mu W$ and $33\mu W$, separately, the smallest MMT can be obtained: ${{\theta }_{\min }}=3.8nrad$. It is clear that increasing the input power can decrease the MMT, meanwhile the linear relationship between the piezoelectric transducer driving voltage and the MMT ${{\theta }_{\min }}$ is shown in Fig.6.\\
\vspace{-0.4cm}
\begin{figure}[htbp]
\includegraphics[width=3.5in]{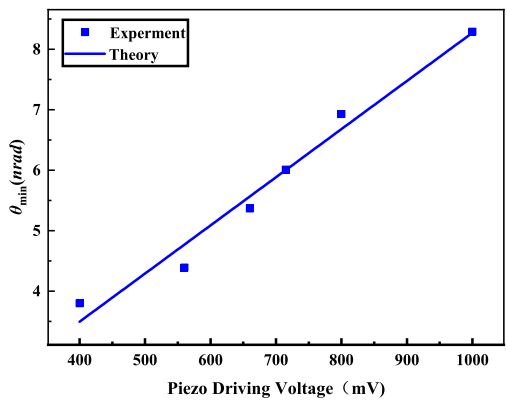}
\caption{\label{fig:epsart} The MMT versus the piezo driving voltage. (The solid line is the theoretical fitting curve.)}
\end{figure}
\vspace{-0.5cm}
\section{Conclusion}
\vspace{-0.2cm}
 We demonstrate the advantage of the combination of WVA technique and BHD system compare with the conventional measurement and the SD system, separately. Then implement the "WVA+BHD" system in an optical small tilt experiment. The weak value measurement based on high-order modes BHD system solves the precision limitation problem that caused by the detector saturation, which is also verified experimentally. Moreover, one can obtain the tilt information of all the photons by detecting only few photons. Our scheme also explains that the destructive interference process is the essence of the weak value measurement, from a classic perspective. The present smallest MMT we obtained is $3.8nrad$. However, by increasing input power and purify the light mode (optimizing the MC), which are not difficult, we can get a more better measurement precision. The WVA+ BHD scheme can be expanded to more application scenarios (extremely weak output, wider measurement bandwidth), and can also achieve faster response time and higher detection efficiency (compared with conventional CCD detection). Based on our experimental setup, if we use the spatially squeezed light, higher measurement precision can be obtained. Furthermore, the precision measurement of the optical small tilt can be valuable and useful in some special quantum precision measurements, such as atomic force microscopy and quantum biological measurement, positioning between satellites and so on.

\vspace{-0.5cm}
\begin{acknowledgments}
\vspace{-0.2cm}
This work was supported by National Key Research and Development Program of China (2021YFC2201802); National Natural Science Foundation of China (NSFC) (11874248, 11874249, 62027821 and 12074233); Natural Science Foundation of Shanxi Province (201801D121007); Research Project Supported by Shanxi Scholarship Council of China (2021-005).
\end{acknowledgments}

\nocite{*}
\bibliography{aipsamp}

\end{document}